\documentclass{pasa}

\jid{PASA}
\doi{10.1017/pas.2015.xxx}
\jyear{2015}

\title[The HMQ Catalogue]{The Half Million Quasars (HMQ) Catalogue}
\author[E. Flesch]{\textbf{Eric W. Flesch }$^{A,B}$\\
\\
\affil{$^A$PO Box 5, Whakatane, New Zealand}
\affil{$^B$Email: eric@flesch.org}}

\begin{document}

\begin{abstract}
A quasar catalogue is presented with a total of 510764 objects including 424748 type 1 QSOs and 26623 type 1 AGN complete from the literature to 25 January 2015.  Also included are 25015 high-confidence SDSS-based photometric quasars with radio/X-ray associations, 1595 BL Lac objects, and 32783 type 2 objects.  Each object is displayed with arcsecond-accurate astrometry, red and blue photometry, redshift, citations, and radio and X-ray associations where present.  Also, 114 new spectroscopically confirmed quasars are presented.
\end{abstract}


\maketitle

\section{Introduction}

We have entered an era of large data where accuracy and completeness are the expected norm in every data-driven endeavour.  If the project includes legacy data, however, then particular challenges are faced to bring that legacy data up to today's standard.  So it is with quasar cataloguing which includes 20$^{th}$ century legacy data having paper-specific astrometry and photometry standards of great variance.  These were catalogued as reported, most recently in the V\'eron-Cetty \& V\'eron Catalogue of Quasars and Active Nuclei 13$^{th}$ edition (VCV: V\'eron-Cetty \& V\'eron 2010), although the V\'erons worked to improve data quality throughout their releases with good success; in their last five releases I contributed a small assist by auditing their pre-release data to find any anomalies.  However, arcsecond-accurate astrometry was not achieved for some thousands of objects and classifications were unreliable for 3\% of objects as will be detailed below.  Thus the legacy quasar data needed to be cleaned to today's standard to integrate that data into the databases of today.

This paper presents a new quasar catalogue which is complete from the literature to 25 January 2015, bears arcsecond-accurate astrometry and uses a higher threshold to accept objects as quasars.  This ``Half Million Quasars'' (HMQ) catalogue was extracted from the on-line unpublished ``Milliquas''\footnote{Milliquas: http://quasars.org/milliquas.htm, also on NASA HEASARC at https://heasarc.gsfc.nasa.gov/W3Browse/all/milliquas.html} catalogue which includes photometric and radio/X-ray associated quasar candidates.  Over the past few years I've worked to resolve all issues remaining in the legacy data to where I can now present 100\% reliable astrometry and 99\% reliable classifications for all objects.  Resolved issues are discussed below and outcomes tabulated.

The HMQ catalogue presents a complete survey of type 1 QSOs/AGN from the literature; type 2 objects are included as a convenience to the user but only at $\approx$90\% completeness from the literature because they are such a heterogeneous group from many criteria used throughout 50 years of discovery; nowadays type 2 is used in a practical way to show narrower emission lines than type 1.  This catalogue signals a shift from the VCV philosophy of including questionable objects (so flagged by their discoverers) as quasars, so such objects are dropped.  BL Lac objects are included to continue VCV's tradition of including them.  Also some high-confidence photometric quasar candidates with radio/X-ray associations are included as discussed in section 8. 

The intent here was to present original quasar names and best redshifts.  VCV usually presented original names but sometimes they substituted other forms based on astrometry or unclear provenance; I recovered some original names along the way.  The SDSS\footnote{Sloan Digital Sky Survey, http://sdss.org} re-surveyed many legacy objects without always acknowledging the original identifications, but those original identifications are reported here.  Part of the problem would have been the discrepant early astrometry but that is now fully resolved.  SDSS redshifts are preferred and usually substituted for original discovery redshifts, and so cited.    
 
Accordingly, the catalogue presented here brings the community's quasar archive to a high standard of usability where each quasar is recorded with arcsecond-accurate astrometry (out to 2 arcsec for some legacy objects), its original name and discovery citation, red and blue photometry, best redshift with citation, and radio/X-ray associations.  The catalogue is available in both flat file and FITS formats\footnote{at http://quasars.org/HMQ.htm} and has a ReadMe which describes the contents and source data therein.  Table 1 shows a few sample lines of the flat file with some explanation of the columns, but the HMQ ReadMe gives full details of the layout and contents.  Figure 1 shows the sky coverage of the HMQ; the high density blocks of coverage are primarily from the SDSS; the high density strip at decl -30$^{\circ}$ is from the 2dF QSO Redshift Survey (2QZ: Croom et al. 2004).

\begin{figure} 
\includegraphics[scale=0.25, angle=0]{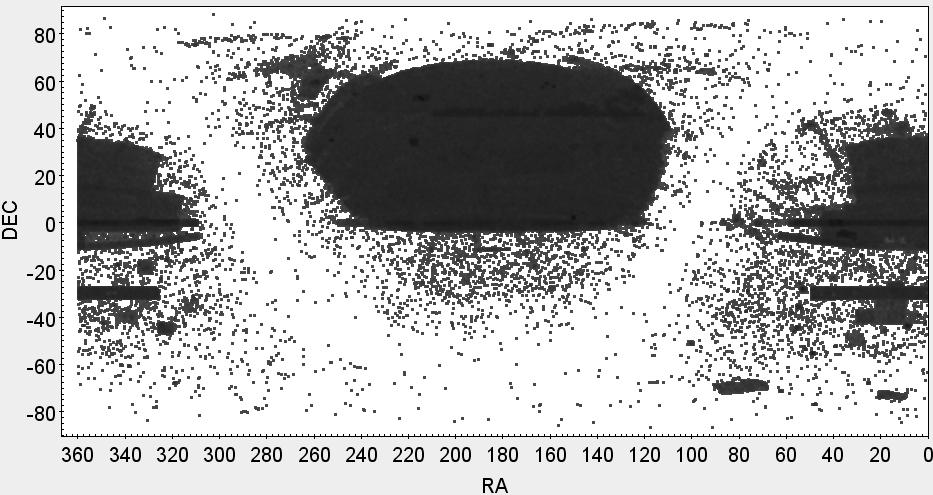} 
\caption{Sky coverage of the HMQ.} 
\end{figure}

\begin{table*} 
\caption{Sample lines from the HMQ.} 
\includegraphics[scale=0.325, angle=0]{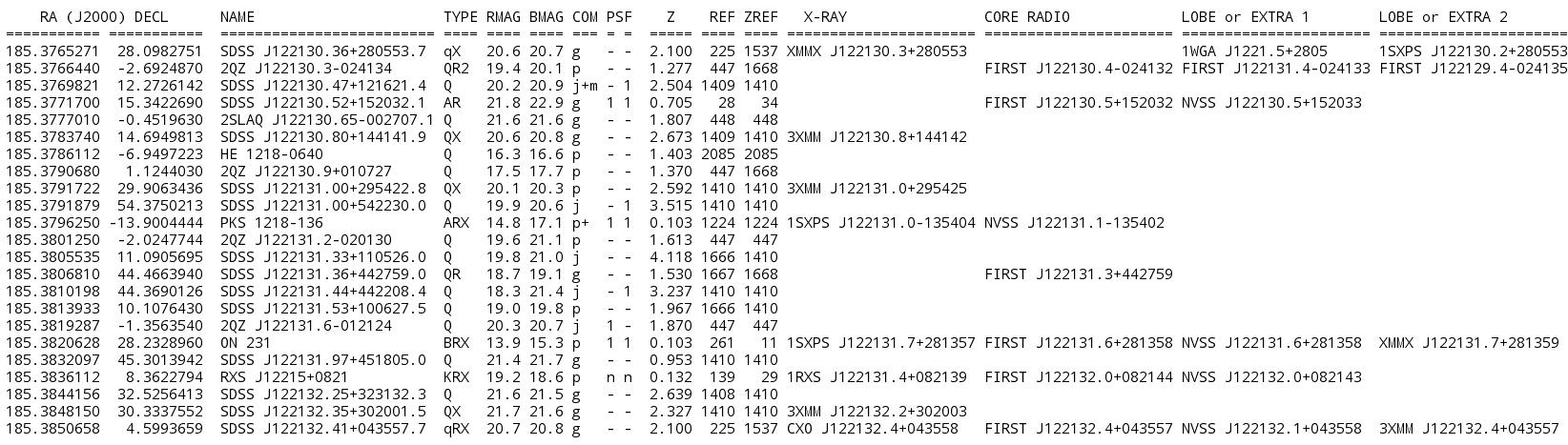} 
\tiny
Notes on columns (see ReadMe for full descriptions): 
\begin{itemize}
	\item TYPE: 1st char is the object classification: Q=QSO, A=AGN, B=BL Lac type, q=photometric, K=type 2; see Section 5.  Extra chars summarize the associations displayed: R=radio, X=X-ray, 2=double radio lobes. 
  \item COM:  comment on photometry: p=POSS-I magnitudes, so blue is POSS-I O, j=SERC Bj, g=SDSS green, +=optically variable, m=nominal proper motion.
  \item PSF:  for red \& blue sources: '-'=stellar, 1=fuzzy, n=no psf available, x=not seen in this colour.
  \item REF \& ZREF: citations for name and redshift; citations are indexed in the file "HMQ-references.txt".
  \item LOBE or EXTRA: if TYPE shows a '2' (=lobes), then double radio lobe identifiers are displayed here.  Otherwise, any additional radio and/or X-ray identifiers are displayed here.
\end{itemize}
The full table can be downloaded from http://quasars.org/HMQ.htm, also available in FITS.
\end{table*}

The sections below describe the issues involved in constructing this catalogue, including data acquisition from the legacy VCV catalogue and the SDSS catalogues which supply most quasars, quasars newly published here, and the inclusion of high-confidence photometric quasars.  AGN classification is detailed.  Photometry and radio/X-ray associations are summarized with references given.

\section{The VCV Optical catalogue}

A fundamental part of constructing the HMQ was to audit and incorporate the best previous comprehensive catalogue of quasars, which is of course the V\'eron-Cetty \& V\'eron Catalogue of Quasars and Active Nuclei 13$^{th}$ edition \cite{VCV}.  However, VCV included astrometry and identifications which were not always optically based, for legacy reasons.  In Flesch \shortcite{FL13} I promulgated the need to convert that to a fully optical presentation to enable easy inclusion into dynamic databases like NED\footnote{NASA/IPAC Extragalactic Database, http://ned.ipac.caltech.edu} and SIMBAD\footnote{SIMBAD database at CDS, http://simbad.u-strasbg.fr/simbad}, and in service of that I presented the true position or status of 449 VCV objects which hitherto had no obvious optical counterpart.  I've continued that task to where the VCV catalogue is now fully mapped, line for line, into the optical J2000.  This ``VCV Optical catalogue''\footnote{the VCV Optical catalogue is downloadable, with a ReadMe, from http://quasars.org/HMQ.htm} identifies and resolves all anomalies including those reported in Flesch \shortcite{FL13}.  Following is a summary of those issues, and lists of objects supplementary to those presented in Flesch \shortcite{FL13}. 

VCV presented 167567 QSOs/AGN of which 160271 were subclassed as type 1 (or had no subclass), 6203 as type 2, 926 as LINER galaxies, and 167 as starburst galaxies; also 1374 BL Lac objects were listed, thus 168941 in total.  The HMQ optical mapping, with the benefit of SDSS data through DR12, resolves these as 156322 type 1, 1142 BL Lac objects, 4662 type 2, 747 LINERs, 5024 galaxies, 267 stars, 667 objects of unknown classification (usually from SDSS-DR7 or later SDSS releases which dropped them), 8 satellite streaks, 2 asteroids (i.e., 3-colour strobes on SDSS), 2 star spikes, 61 duplicates of other VCV objects, and 37 objects which are lost or never existed.  It behoves me to explain these reclassifications, so first I list out the most involved discovery papers (the individual objects involved are listed in the VCV Optical catalogue with these citations): \\    
   
\noindent (A) 4113 objects were re-classified as stars or non-AGN galaxies by papers subsequent to VCV, usually SDSS releases:
\begin{enumerate}
  \item 3424 by SDSS-DR8 \cite{DR8} due to re-processing of the spectra via an improved algorithm. 
  \item 291 by SDSS-DR12Q \cite{DR12Q} which reclassified many objects via visual inspection of the spectra.
  \item 267 by SDSS-DR9Q \cite{DR9Q} which reclassified many objects via visual inspection of the spectra.
  \item 42 by SDSS-DR10Q \cite{DR10Q} which reclassified many objects via visual inspection of the spectra.
  \item 39 by SDSS-DR9 \cite{DR9} via an improved automated pipeline.
  \item 25 by SDSS-DR10 \cite{DR10} via an improved automated pipeline.
  \item 14 by Kleinman et al. \shortcite{KLEINMAN} which identified white dwarfs from the SDSS-DR7 data.
  \item 5 by Ge et al. \shortcite{GE} which identified double-peaked narrow emission line galaxies (NELGs).
  \item and 6 other objects from 4 papers.
\end{enumerate}
   
\noindent (B) 1845 objects were originally classified as stars or non-AGN galaxies by their discovery papers but taken up by VCV as ``active galaxies'', or were reclassified as non-AGN prior to VCV's final edition but VCV retained the original classification: 
\begin{enumerate}
  \item 473 from Schneider/Schmidt/Gunn \shortcite{SSG94} which classified these as emission-line galaxies (ELGs) of z$<$0.45.
  \item 430 from SDSS-DR7 \cite{DR7} which classified these as stars or non-AGN galaxies, confirmed by their exclusion from SDSS-DR7Q \cite{DR7Q}.  Many of these had been classified as AGN by earlier publications.   
  \item 261 from the 2dF Galaxy Redshift Survey (2dF-GRS: Colless et al. 2001) which is a large galaxy catalogue without QSO classifications.  VCV extracted high-redshift objects and others with emission lines showing activity, but visual inspection of the high-redshift objects reveal that they are often confused with closely co-positioned neighbours.  Some of these objects were cited elsewhere and included by VCV on that basis, but subsequent SDSS classifications are as galaxies only.  The optical appearance of almost all of these objects is that of galaxies, and only 16 show radio/X-ray association.  I see no avenue by which to retain these objects as type I AGN, so they are dropped.
  \item 144 from Iovino/Clowes/Shaver \shortcite{IOVINO} were annotated by them with '?' signifying spectral overlap or possible star.  Many of these are anomalously bright and none show radio/X-ray association, so they are all dropped.  Some of the remaining 1140 objects from that paper are quite bright or reddish, so caution in their use is indicated. 
  \item 88 from the 2QZ \cite{CROOM04}, which classified 85 of these as ``continuum'' objects and 3 as NELGs.  None show radio/X-ray association and all of the continuum objects were marked by VCV as ``BL?'' showing that they were questionable, so they are dropped. 
  \item 64 from Mauch \& Sadler \shortcite{MAUCH} who classified these as LINERs.
  \item 47 from Schneider/Schmidt/Gunn \shortcite{SSG99} who, as with their 1994 paper, classified these as ELGs of z$<$0.45.  
  \item 45 from La Franca et al. \shortcite{LF} which classified these as ELGs.
  \item 25 from Barcons et al. \shortcite{BARCONS} which classified these as NELGs. 
  \item 18 from Sandage \& Bedke \shortcite{SB} which classified these NGC galaxies as LINERs and starburst. 
  \item 10 from Gaston \shortcite{BGA} and 2 from Gaston \shortcite{BGB}, both of which surveyed SA57 (i.e., `selected area 57'), but these 12 are without quasar colours or radio/X-ray association.  Gaston \shortcite{BGB}, sec III, qualified that the SA57 photometry had ``much greater scatter'' and ``nonuniformity'' compared with concomitant CFHT and KPNO photometry, and in sec II\textit{ai} stated a forthcoming paper, ``Hoag \textit{et al.} 1983'', of these SA57 quasars which however never appeared.  Edwards/Beauchemin/Borra \shortcite{EBB} resurveyed SA57, confirming no Gaston object, and related that Hoag in fact ``did not confirm many of the Gaston candidates''.
  \item and 238 other objects from 117 papers, usually from VCV's own annotations that they are questionable, or starburst galaxies, etc.  
\end{enumerate}
   
\noindent A total of 61 duplications are found in VCV, of which 39 were presented in Flesch \shortcite{FL13} and one in Cupani et al. \shortcite{CUPANI}.  The remaining 21 newly-found duplications are listed in Table 2 which displays the VCV names of the wrongly-sited duplicate and correctly-sited master, the offset between them in arcseconds, and the final HMQ name which is usually the original form.  One of these, \#10 on the list, deserves a special mention in that VCV took the redshift of the master from SDSS-DR6, but that redshift was corrected by SDSS-DR7 to that given by the original discoverers. 
    
As a further supplement to Flesch \shortcite{FL13}, Table 3 lists 31 more moves of 30+ arcseconds for VCV objects.  29 of these are newly presented, whilst two, \#s 4 \& 5, are moved from the position given in Flesch \shortcite{FL13} due to better information.  Also, four objects were wrongly moved in Flesch (2013), being Q 0004-019, IRAS 12202+1646, and MS 22225+2114 which are returned to their VCV location, and IXO 32 which is moved to its finding chart location at 09 10 20.2 +07 05 55 (J2000).  The right-hand column is a comment on the move, ``OA''=original author.   

Also, Flesch \shortcite{FL13} identified 30 VCV objects to de-list because they either are not quasars or have information of such poor quality that the object is lost beyond any mechanism of recovery.  In a similar vein I present 14 VCV objects in Table 4 which are not seen at all, and the methods of Flesch \shortcite{FL13} find no candidate.  These objects are obviously not included in the HMQ catalogue.

HMQ has sourced much data directly from large source catalogues like SDSS, 2QZ \cite{CROOM04}, etc., so the actual uptake from VCV consists of a residue from smaller papers which amount to 19127 type 1 objects, 1083 BL Lac objects, and 1594 type 2 objects for a total of 21794 legacy objects.  Most of the citations for these legacy objects have been copied from the VCV citation list.

\begin{table*}
\scriptsize	 
\caption{21 more duplicates in VCV13}
\tiny
\begin{tabular}{r@{\hspace{4pt}}|l@{\hspace{4pt}}c@{\hspace{4pt}}c@{\hspace{4pt}}c@{\hspace{4pt}}|r@{\hspace{4pt}}|l@{\hspace{4pt}}
                c@{\hspace{4pt}}c@{\hspace{4pt}}c@{\hspace{4pt}}|l@{\hspace{4pt}}c@{\hspace{4pt}}c@{\hspace{4pt}}c@{\hspace{4pt}}}
\hline 
   & \multicolumn{4}{c}{VCV Duplicate} & offset & \multicolumn{4}{c}{VCV Master} & \multicolumn{4}{c}{HMQ} \\
\# & Name & tbl$^{*}$ & z & Vmag & arcsec & Name & tbl$^{*}$ & z & Vmag & Name & z & Rmag & Bmag \\
\hline
 1 & IGR J00256+6821   & 3  & 0.012  &       & 665 E/W & IRAS 00227+6805  & 3  & 0.012  &       & ZOAG 120.54+5.61 & 0.012  & 13.5  & 18.3 \\
 2 & Q 0235+164        & 3  & 0.524  & 20.9  & 2       & AO 0235+164      & 1  & 0.940  & 15.5  & Q 0235+164  & 0.940  & 18.8  & 20.2 \\
 3 & Q 03020-0014      & 1  & 3.050  & 21.5  & 26      & SDSS J03045-0002 & 1  & 3.061  & 20.5  & Q 03020-0014  & 3.060  & 20.3  & 20.5 \\
 4 & 1WGA J0827.6+0942 & 1  & 0.260  & 15.5  & 14      & SDSS J08276+0942 & 3  & 0.260  & 18.8  & SDSS J082739.00+094209.4  & 0.260  & 17.3  & 18.7 \\
 5 & JGR J09253+6929   & 3  & 0.039  &       & 63      & IGR J09253+6929  & 3  & 0.039  &       & IGR J09253+6929  & 0.039  & 13.4  & 15.2 \\
 6 & RX J09278+3431    & 1  & 0.425  & 17.8  & 37 E/W  & SDSS J09278+3431 & 3  & 0.426  & 19.5  & SDSS J092751.11+343103.6  & 0.426  & 17.8  & 18.1 \\
 7 & 1WGA J0958.0+4903 & 3  & 0.234  & 18.7  & 14      & SDSS J09580+4903 & 3  & 0.242  & 19.0  & 1WGA J0958.0+4903  & 0.242  & 17.5  & 18.2 \\
 8 & RX J10514+3358    & 3  & 0.183  & 18.3  & 48      & SDSS J10514+3358 & 3  & 0.183  & 19.4  & SDSS J105128.41+335850.8  & 0.183  & 17.1  & 19.0 \\
 9 & LH 28B            & 3  & 0.205  & 18.8  & 734 E/W & RX J10543+5725   & 3  & 0.205  & 18.2  & RX J10543+5725  & 0.205  & 18.4  & 21.3 \\
10 & FIRST J10544+2534 & 1  & 2.400  & 17.5  & 120 N/S & SDSS J10544+2536 & 1  & 1.774  & 18.4  & FBQS J105427.1+253600  & 2.400  & 17.5  & 18.8 \\
11 & SBS 1059+599      & 1  & 1.830  & 17.6  & 47      & TEX 1059+599     & 1  & 1.830  & 17.1  & TEX 1059+599  & 1.830  & 16.9  & 17.5 \\
12 & WEE 43            & 1  & 3.009  & 19.8  & 21      & SDSS J11161+1358 & 1  & 3.034  & 20.0  & WEE 43  & 3.033  & 19.4  & 20.9 \\
13 & RXS J11479+2715   & 3  & 0.364  & 19.6  & 60 E/W  & US 2964          & 1  & 0.363  & 16.1  & US 2964  & 0.363  & 16.3  & 16.4 \\
14 & Q 1207+106        & 1  & 2.330  &       & 150 E/W & SDSS J12098+1021 & 1  & 2.312  & 19.4  & Q 1207+106  & 2.313  & 19.3  & 19.8 \\
15 & SDSS J13523+1302  & 1  & 1.600  & 18.1  & 0       & H 1400\#16A       & 1  & 1.600  & 18.1  & H 1400\#16A  & 1.600  & 17.7  & 18.0 \\
16 & 3C 295.0          & 3  & 0.461  & 19.8  & 60 N/S  & SDSS J14113+5212 & 3  & 0.462  & 20.8  & 3C 295.0  & 0.461  & 18.0  & 20.8 \\
17 & RXS J15244+3032   & 1  & 0.282  & 17.2  & 6       & SDSS J15244+3032 & 1  & 0.274  & 16.9  & SDSS J152428.67+303237.5  & 0.274  & 16.6  & 16.7 \\
18 & Q 1532+01/2       & 1  & 0.310  & 17.5  & 90 E/W  & XBS J15349+0130  & 3  & 0.310  & 18.8  & SDSS J153456.20+013032.3  & 0.308  & 17.2  & 18.4 \\
19 & ESO 136-G06       & 3  & 0.015  & 14.8  & 43056   & ESO 136-G006     & 3  & 0.015  & 14.8  & ESO 136-6  & 0.015  & 13.4  &   \\
20 & ESO 407-IG17      & 3  & 0.091  & 17.4  & 16      & NVSS J23243-3426 & 3  & 0.091  & 16.1  & ESO 407-17  & 0.091  & 14.5  & 14.9 \\
21 & CXOMP J23485+0053 & 1  & 0.769  & 20.0  & 19      & CXO J23485+0058  & 3  & 0.948  & 21.3  & CXO J23485+0058  & 0.948  & 18.8  & 19.3 \\
\hline
\multicolumn{12}{l}{$^{*}$ VCV table 1 listed QSOs, table 3 listed AGN.} \\
\end{tabular}
\end{table*}

\begin{table*}
\scriptsize	 
\caption{31 more moves of 30+ arcseconds in VCV13}
\tiny
\begin{tabular}{r@{\hspace{4pt}}|l@{\hspace{2pt}}l@{\hspace{4pt}}c@{\hspace{4pt}}c@{\hspace{4pt}}|
                r@{\hspace{2pt}}|l@{\hspace{2pt}}l@{\hspace{4pt}}|l}
\hline 
   & \multicolumn{4}{c}{VCV} & move & \multicolumn{2}{c}{HMQ} \\
\# & Name & J2000 & tbl$^{*}$ & z & asec & Name & J2000 & comment \\
\hline
 1 & ChOPS J04206+3245  & 042348.6+324522  & 3  & 0.307  & 2271 & ChOPS J04206+3245 & 042048.6+324522  & offset 3 time minutes E/W, VCV error \\
 2 & ESO 374-G25  & 100323.6-373339  & 3  & 0.023  & 594  & ESO 374-25  & 100323.7-372345  & offset 10 arcmin N/S \\
 3 & Q 1257+2843  & 125951.2+282730  & 1  & 2.990  & 589  & Q 1257+2843  & 125951.1+281741  & offset 10 arcmin N/S, OA error \\
 4 & Q 0112-27  & 011452.8-271409  & 1  & 2.894  & 343  & Q 0112-27  & 011517.2-271223  & correct position courtesy of Cyril Hazard \\
 5 & Q 0752+617  & 075622.4+613401  & 1  & 1.892  & 234  & Q 0752+617  & 075646.6+613639  & optical fit like co-published Q 0932+501 \\
 6 & RX J08087+0018  & 080847.1+001801  & 3  & 0.019  & 124  & UGC 4248  & 080855.4+001806  & corrected to galaxy position \\
 7 & MCG -03.27.026  & 104217.6-174018  & 3  & 0.021  & 85  & MCG -3-27-026  & 104218.9-173855  & corrected to galaxy position \\
 8 & F 265  & 065617.3-653348  & 3  & 0.029  & 78  & FAIRALL 265  & 065629.8-653338  & corrected to galaxy position \\
 9 & EXO 1429.9+3717  & 143158.6+370402  & 1  & 1.925  & 74  & EXO 1429.9+3717  & 143201.7+370258  & within OA error circle, NBCKDE J143201.72+370257.8 \\
10 & F 917  & 205255.5-515246  & 3  & 0.053  & 64  & ESO 235-10  & 205259.5-515154  & corrected to galaxy position \\
11 & G 2344-3852  & 234649.2-383520  & 3  & 0.041  & 63  & PGC 2817464  & 234643.8-383521  & offset 1 arcmin E/W \\
12 & ESO 353-G09  & 013150.4-330809  & 3  & 0.017  & 59  & ESO 353-9  & 013150.4-330710  & offset 1 arcmin N/S \\
13 & Q 1409+732  & 140949.9+725941  & 1  & 3.560  & 58  & Q 1409+732  & 141003.2+725939  & FC was onto plate flaw, 1 arcmin E/W \\
14 & ESO 338-G17  & 193813.1-385425  & 3  & 0.019  & 58  & ESO 338-17  & 193816.4-385342  & corrected to galaxy position \\
15 & MCG -03.04.046  & 011925.0-154107  & 3  & 0.051  & 53  & MCG -3-04-046  & 011925.0-154200  & offset N/S \\
16 & MCG -01.02.013  & 002600.0-025507  & 3  & 0.068  & 51  & MCG -1-02-013  & 002600.0-025558  & offset 50 arcsec N/S \\
17 & MCG -03.02.027  & 003818.4-145007  & 3  & 0.037  & 49  & MCG -3-02-027  & 003818.4-145056  & offset 50 arcsec N/S \\
18 & Q 0139-371  & 014158.1-365251  & 1  & 0.365  & 48  & Q 0139-371  & 014201.8-365310  & finding chart of spectra, OA \\
19 & MCG -01.25.049  & 095056.5-045906  & 3  & 0.022  & 47  & MCG -1-25-049  & 095056.4-045953  & offset 45 arcsec N/S \\
20 & MCG -01.40.001  & 153318.9-084125  & 3  & 0.023  & 45  & MCG -1-40-001  & 153320.6-084203  & corrected to galaxy position \\
21 & F 239  & 044851.2-573937  & 3  & 0.023  & 43  & ESO 119-8  & 044856.6-573935  & corrected to galaxy position \\
22 & MCG -01.30.005  & 113423.3-094006  & 3  & 0.021  & 42  & NGC 3722  & 113423.2-094048  & corrected to galaxy position \\
23 & MCG -02.27.009  & 103527.3-140707  & 3  & 0.015  & 41  & MCG -2-27-009  & 103527.3-140748  & offset 40 arcsec N/S \\
24 & ESO 104-041  & 190933.3-622839  & 3  & 0.085  & 41  & ESO 104-41  & 190938.8-622855  & E gal is AGN, XMMSL J190938.8-622855 \\
25 & MCG -02.08.039  & 030028.2-112504  & 3  & 0.030  & 37  & MCG -2-08-039  & 030030.6-112456  & corrected to galaxy position \\
26 & G J0639-5125  & 063936.8-512515  & 3  & 0.108  & 37  & DENISP\_G\_J0639411-512515  & 063940.7-512515  & offset 4 timesec E/W \\
27 & Q 21570-302   & 215959.3-300027  & 1  & 1.900  & 36  & Q 21570-302 & 215956.6-300027  & 35 arcsec offset E/W \\
28 & MCG -02.11.014  & 040427.5-101006  & 3  & 0.031  & 35  & MCG -2-11-014  & 040427.5-101041  & moved 35 arcsec S \\
29 & F 296  & 022504.3-631320  & 3  & 0.057  & 32  & F 296  & 022509.0-631327  & moved 5 timesec E \\
30 & 2E 1416+2523  & 141858.1+251020  & 1  & 0.674  & 32  & 2E 1416+2523  & 141857.6+250949  & OA specified wrong object, SDSS-DR10 re-found. \\
31 & Q 1606.5+29.1  & 160831.1+290231  & 3  & 0.060  & 32  & KP 1606.5+29.1  & 160831.6+290200  & OA finding chart \\
\hline
\multicolumn{9}{l}{$^{*}$ VCV table 1 listed QSOs, table 3 listed AGN.} \\
\end{tabular}
\end{table*}

\begin{table*}
\scriptsize	 
\caption{14 more objects to delete from VCV13}
\tiny
\begin{tabular}{rlllrrl}
\hline 
\# & Name & J2000 & tbl$^{*}$ & mag$^{\dagger}$ & z & comment \\
\hline
 1 & Q J02397-0135  & 023947.5-013512  & 1  &    & 1.605  & not seen \\
 2 & XMM 551B  & 031243.0-551157  & 1  &    & 0.789  & not seen \\
 3 & RXS J04014-0801  & 040126.3-080200  & 1  & 14.6  & 0.147  & not seen at v=14.6, OA says "QSO?" questionable \\
 4 & Q 0537-44 A1  & 053850.6-440508  & 3  & 21.0  & 0.885  & unseen binary, OA states unclear \\
 5 & 1H 0828-706  & 082817.2-704859  & 1  & 16.7  & 0.239  & approx position, no candidate seen \\
 6 & RXS J09052-3350  & 090517.9-335016  & 3  & 20.5  & 0.425  & not seen, OA says NELG \\
 7 & COSM J09591+0228  & 095907.7+022816  & 1  & 18.6  & 0.800  & not seen \\
 8 & 2QZ J095926+0115B  & 095926.0+011546  & 1  & 20.3  & 1.404  & not seen, OA presents no double \\
 9 & EXO 1145.1+0031  & 114741.2+001455  & 3  & 19.2  & 0.096  & not seen \\
10 & Q J13053+2902  & 130520.1+290247  & 3  & 20.2  & 0.076  & multiple candidates, unclear if any is QSO \\
11 & Q J13061+2850  & 130609.1+285028  & 3  & 22.1  & 0.079  & not seen \\
12 & SDSS J14111+4249  & 141107.0+424936  & 1  &    & 1.075  & SDSS shows nothing there \\
13 & 284\_4709  & 141630.7+520407  & 3  & 19.8  & 0.450  & nothing seen \\
14 & RXS J21527+1341  & 215242.4+134150  & 1  & 16.2  & 0.202  & cluster, unclear, OA does not state QSO \\
\hline
\multicolumn{7}{l}{$^{*}$ VCV table 1 listed QSOs, table 3 listed AGN.} \\
\multicolumn{7}{l}{$^{\dagger}$ Vmag from VCV, can be blank.} \\
\end{tabular}
\end{table*}

\section{Inclusion of SDSS quasars}

The majority of quasars in this catalogue are, of course, sourced from the ongoing SDSS releases which have dwarfed in size and complexity all previous quasar surveys; a full description of the SDSS project is beyond the scope of this paper, so the reader is assumed to have some acquaintance with it.  There are two simplifications in the HMQ citations: the SDSS data releases 1 and 2 are subsumed into the DR3 release, consistent with their presentation in the SDSS data archives, and DR11 which was an internal release only is similarly subsumed into DR12 and so not considered further.    

In use of the SDSS, it's not enough to simply take up all SDSS ``quasars'' because they aren't as uniformly reliable as that.  The SDSS project uses an automated pipeline which matches spectra to standard templates by which any spectrum well-matched to a quasar template is securely classified as a quasar.  But of course many spectra are ambiguous fits to templates, especially for faint objects with noisy spectra.  To cater for this, the SDSS consortium periodically issues manually-checked quasar catalogues which are much more reliable than the automated ones, the most recent being SDSS-DR12Q \cite{DR12Q}.  I use those manual catalogues as a definitive source although they do have a few false objects to remove -- indeed, each edition typically drops some objects from previous editions.  Objects not accepted into those manual catalogues (i.e., ``pipeline-only'' objects) may not be quasars or their spectra could be too low quality (i.e.,``low significance'') to be credible, but often they simply fell on the wrong side of threshold cutoffs such as magnitude or luminosity.  Inspection of the spectra of such pipeline-only quasars shows that indeed many are plainly recognizable as quasar spectra, and many are not.  Therefore I needed to find criteria to bulk-separate the high significance pipeline-only quasars from the low, in order to accept the good ones into this comprehensive catalogue.  Fortunately this could be done as follows. 

As a background, the SDSS data releases 1 to 7 comprise the ``legacy'' period in which the classifications of quasars, galaxies, stars, and ``unknowns'' were released as separate incremental tables, and the manual catalogues were quasars only.  There was no means of bulk-grading the pipeline-only quasars in the legacy data.  The current ``SDSS3'' data releases 8 to 12 are made as single tables which include all classifications and historical data, and the manual catalogues include a ``superset'' of all manually inspected classifications.  Key enhancements of the SDSS3 releases are the provision of a ZWARNING field which flags problems with the spectra, and a SUBCLASS field which displays values like ``BROADLINE'' or ``STARFORMING'', but is often blank.  For the ZWARNING field, I follow documented recommendations and accept only objects with ZWARNING of 0 (no warnings) or 16 (saturation) as well as 256 (poor astrometry for plughole) and some 128 (unplugged, but flag is documented to be wrong for about 200 objects which have strong signal).  For the SUBCLASS field, I find from inspection of hundreds of spectra that the better quality spectra bear a populated SUBCLASS field, and the pipeline classifications and redshifts for those objects are of a good significance.  In contrast, the spectra without a subclass are usually noisy and the pipeline information of a low significance -- which undoubtedly explains the pipeline's inability to discern a subclass.  Accordingly, I take a populated SUBCLASS field to signal a high significance spectrum. 

Thus, by use of the SUBCLASS and ZWARNING fields I am able to bulk-separate the pipeline-only quasars into high significance and low significance spectra, thus enabling inclusion of those high significance quasars into this catalogue.  An additional benefit is that there is a high overlap of the SDSS3 data with the legacy SDSS data; thus all legacy pipeline-only quasars can be dropped unless qualifying by this SDSS3-based rule.  However, pipeline-only quasars are also kept if confidently associated to radio/X-ray detections, even if the spectrum is of low significance.  Also, pipeline-only redshifts above z=4 are excluded as per advice from Adam Myers; upon further inspection of high-z pipeline-only spectra, I've lowered this cutoff to z=3.65.      

The total counts of SDSS-discovered type-1 objects included into HMQ are 369249 manually checked quasars and 20298 pipeline-only quasars.  Also, 11821 SDSS redshifts have been applied to quasars from earlier papers, the SDSS redshifts being favoured due to their precision and reliability.  Specific breakdowns are available from the HMQ data in which each object is referenced with its discovery citation and redshift citation.  Table 5 lists the discovery papers which contribute most quasars to the HMQ catalogue, in numerate order; there it is seen that the SDSS releases dominate the quasar portfolio today.  A note on the table 5 column ``\# of type 1'': this shows how many objects were originally presented by that paper, but SDSS astrometrically-based object names can change slightly between releases; the HMQ presents the current names of SDSS objects, so the discovery citation is that of the SDSS release which first presented the current name.  My processing showed that DR12Q added 121173 new type-1 quasars, so the 154958 displayed in table 5 includes 33785 quasars with small name changes from earlier releases.

\begin{table}[t] 
\scriptsize	 
\caption{Top 25 discovery papers for the HMQ}
\tiny
\begin{tabular}{@{\hspace{0pt}}r@{\hspace{4pt}}l@{\hspace{2pt}}r@{\hspace{2pt}}r@{\hspace{4pt}}l}
\hline 
\  &    & \# of  & \# of \\
\# & ID & type 1 & redshifts & paper \\
\hline
 1 & SDSS DR12Q manual & 154958 & 298959 & P\^{a}ris et al. \shortcite{DR12Q} \\ 
 2 & SDSS DR9Q manual  & 62478  & 218    & P\^{a}ris et al. \shortcite{DR9Q} \\
 3 & SDSS DR10Q manual & 58006  & 808    & P\^{a}ris et al. \shortcite{DR10Q} \\
 4 & SDSS DR3Q$^{*}$ manual & 39247  & 9      & Schneider et al. \shortcite{DR3Q} \\
 5 & SDSS DR5Q manual  & 28259  & 73     & Schneider et al. \shortcite{DR5Q} \\
 6 & SDSS DR7Q manual  & 26301  & 80222  & Schneider et al. \shortcite{DR7Q} \\
 7 & XDQSO candidates  & 23135  & 0      & Bovy et al. \shortcite{XDQSO} \\
 8 & 2QZ/6QZ           & 23036  & 19692  & Croom et al. \shortcite{CROOM04} \\
 9 & 2SLAQ             & 7722   & 6556   & Croom et al. \shortcite{CROOM09} \\
10 & SDSS DR12 auto    & 4261   & 5781   & Alam et al. \shortcite{DR12} \\
11 & PGC$^{\dagger}$   & 4254   & 5      & Paturel et al. \shortcite{PGC} \\
12 & SDSS DR8 auto     & 3417   & 1762   & Aihara et al. \shortcite{DR8} \\
13 & SDSS DR9 auto     & 2952   & 2770   & Ahn et al. \shortcite{DR9} \\
14 & SDSS DR3* auto    & 2597   & 1927   & Abazajian et al. \shortcite{DR3} \\
15 & SDSS DR10 auto    & 2591   & 2300   & Ahn et al. \shortcite{DR10} \\
16 & NBCKDE candidates & 1878   & 14205  & Richards et al. \shortcite{NBCKDE} \\
17 & SDSS DR7 auto     & 1759   & 2507   & Abazajian et al. \shortcite{DR7} \\
18 & DEEP2 Redshifts   & 1437   & 1400   & Newman et al. \shortcite{NEWMAN} \\
19 &                   & 1140   & 1141   & Iovino/Clowes/Shaver \shortcite{IOVINO} \\
20 & SDSS DR6 auto     & 1043   & 1487   & Adelman-McCarthy et al. \shortcite{DR6} \\
21 & SDSS DR4 auto     & 855    & 1237   & Adelman-McCarthy et al. \shortcite{DR4} \\
22 & AGES survey       & 848    & 848    & Kochanek et al. \shortcite{AGES} \\
23 & SDSS DR5 auto     & 823    & 1308   & Adelman-McCarthy et al. \shortcite{DR5} \\
24 & Magellanic Quasars & 543   & 574    & Kozlowski et al. \shortcite{KOZLOWSKI} \\
25 & LAMOST Andromeda  & 519    & 455    & Huo et al. \shortcite{HUO} \\
\hline
\multicolumn{5}{l}{$^{*}$ DR3Q and DR3 here include the earlier DR1 and DR2 releases.} \\
\multicolumn{5}{l}{$^{\dagger}$ The Principal Galaxy Catalogue (PGC) is not actually a discovery} \\
   & \multicolumn{4}{l}{paper, but is used as a reference for names of AGN galaxies.} \\
\end{tabular}
\end{table}

\section{New Quasars}

114 new quasars with spectral confirmation are presented in this catalogue.  They are 92 quasars from Cyril Hazard's work in the 1980s, 4 quasars from E. Margaret Burbidge (2003), and 18 quasars from Dan Weedman on a re-inspection of Weedman \shortcite{DW85}.  In detail: \\  
\\
(1) 92 quasars from Cyril Hazard, the prominent early quasar researcher who identified many quasars from objective prism plates obtained with the UK Schmidt telescope in the 1980s \cite{HAZ80}.  Hazard's \textit{modus operandi} was to share his quasars out with other researchers rather than publish them himself.  As a consequence, many remained unpublished.  In June 2013 he sent word that he is preparing a comprehensive presentation of his works, and we shared some data.  These 92 new quasars were identified from that data, and Cyril graciously assented to their inclusion in this work.  The quasar names are from his worksheets which I have prefaced with ``HAZ'' to clarify their provenance; however, two objects had the same worksheet identifier of ``122'' so I have renamed one as simply ``HAZARD''.  24 of these objects have been covered by the subsequent SDSS DR12 release with all confirmed as quasars -- these are readily identified in the catalogue by the redshift citation.  A sample is shown in Table 6, and the complete set can be extracted from the catalogue by selecting on names beginning with ``HAZ'', or by the citation of Hazard \shortcite{HAZ13}.  Full information on these quasars and a larger set of work is in preparation by Cyril Hazard as of his last communication in September 2013.  

\begin{table}[h]
\scriptsize	 
\caption{Sample quasars from Cyril Hazard}
\begin{tabular}{l@{\hspace{4pt}}l@{\hspace{4pt}}r@{\hspace{4pt}}r@{\hspace{8pt}}r}
\hline 
   Name & J2000 & Rmag & Bmag & z \\
\hline
   HAZ 878  & 00 43 47.6 -26 54 16  & 18.5  & 18.7  & 1.290 \\
   HAZ R157  & 00 45 04.9 -21 53 01  & 18.2  & 18.0  & 2.766 \\
   HAZ 1113  & 00 46 35.3 -26 42 55  & 19.1  & 19.4  & 2.160 \\
   HAZ 0040+6  & 00 48 34.2 +07 53 44  & 18.6  & 18.9  & 2.800 \\
   HAZ R077  & 00 50 59.9 -24 11 38  & 18.2  & 18.7  & 2.958 \\
   HAZ 0120-3G  & 01 25 00.2 +04 01 48  & 18.6  & 19.6  & 2.100 \\
   HAZ 0128+01  & 01 31 18.9 +01 24 36  & 17.9  & 19.4  & 3.094 \\
   HAZ 62  & 04 49 03.2 -17 18 06  & 17.9  & 19.2  & 3.060 \\
   HAZ 12-5-43  & 12 09 35.2 +09 52 01  & 19.5  & 20.0  & 2.069 \\
   HAZ A136.1  & 14 58 45.2 +08 29 49  & 19.0  & 19.6  & 2.647 \\
\hline
\end{tabular}
\end{table}

\begin{table*}
\scriptsize	 
\caption{Quasars from E. Margaret Burbidge \shortcite{EMB}}
\begin{tabular}{rllrrrl}
\hline 
   no. & Name & J2000 & Rmag & Bmag & z & X-ray \\
\hline
   1 & CXO J02224+4221 & 02 22 24.4 +42 21 38 & 17.8 & 18.1 & 1.18 & CXO J022224.4+422138 \\
   2 & EMBUR 2 & 02 22 25.3 +42 24 50 & 20.3 & 21.0 & 0.55 & CXO J022225.3+422450 \\
   3 & EMBUR 3 & 02 22 32.6 +42 26 19 & 19.9 & 21.3 & 0.68 & CXO J022232.6+422619 \\
   4 & EMBUR 4 & 02 22 13.9 +42 25 16 & 19.5 & 20.3 & 2.17 & 3XMM J022213.9+422517 \\
   5 & EMBUR 5 & 02 22 53.1 +42 13 06 & 19.5 & 19.6 & 2.27 & 3XMM J022253.1+421308 \\
\hline
\end{tabular}
\end{table*}

\noindent (2) Four quasars from E. Margaret Burbidge, observed on the Keck-I LRIS on 2 October 2003.  Six targets were selected via X-ray association in the vicinity of NGC 891, and five found to be quasars; the sixth (J022302.5+421731) was missed because the slit was mislaid onto two adjacent stars.  Burbidge elected not to publish due to the sparse results, but released the material to associates, myself included.  The 5 quasars are listed in Table 7 with X-ray associations.  The spectra were reduced by C. Guti\'{e}rrez using IRAF\footnote{IRAF is the Image Reduction and Analysis Facility, http://iraf.noao.edu} and are displayed in Figure 2 which I have annotated with line identifications.  The first quasar has since been re-surveyed by the Serendipitous Extragalactic X-ray Source Identification program (SEXSI: Eckart et al. 2006), so the remaining four are newly published here.  I've named them as EMBUR 2 (etc.) after the discoverer, and their citation is Burbidge \shortcite{EMB}. \\
\\
(3) 18 quasars with probable redshifts from Dan Weedman.  These were first presented in Weedman (1985) but without redshifts because each showed only a single emission line which could not be securely identified as Ly$\alpha$ or CIII, etc.  Today we have SDSS-based photometric redshifts for these 18, primarily from Richards et al. \shortcite{NBCKDE}, which have enabled unambiguous identifications of each single emission line on the premise that the photometric redshift is near to the true redshift.  These quasars have names of WEE 35, WEE 49, etc., from the original paper, and have a redshift citation of Weedman \shortcite{DW14} except for three re-surveyed by DR12 which confirm these as quasars.

\begin{figure} 
\begin{center} 
\includegraphics[scale=0.4, angle=0]{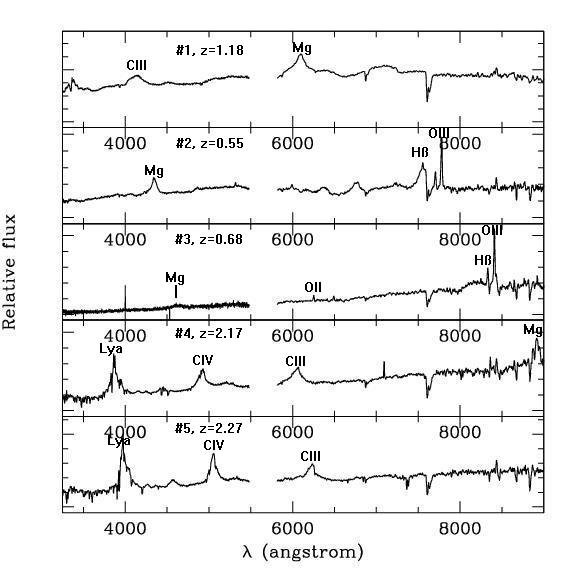} 
\caption{Spectra of the five Burbidge quasars, obtained with Keck-I LRIS using a 5600\AA\ dichroic.} 
\end{center} 
\end{figure}

\section{Classification of the presented objects}

There are 5 object classes reported in this catalogue which are shown with their counts in the first two columns of Table 8.  Type 1 objects (class='Q' \& 'A') are the main thrust of this catalogue and only high confidence objects with secure redshifts are included; where the discovery paper flags objects as questionable, those are dropped.  SDSS quasars are included when presented by the manual surveys such as DR12Q, and also good confidence pipeline-only quasars are accepted as detailed in Section 3.  However, the SDSS pipeline uses a ``BROADLINE'' subclassifier for broad-line emitters, so emitters not flagged as ``BROADLINE'' are either narrow-line emitters or the spectrum was too low confidence to make a judgement.  Therefore to include pipeline-only quasars as type-1 objects I require them to be either of good confidence with the ``BROADLINE'' subclass, or of low confidence (i.e., blank subclass) with secure radio/X-ray association.  

Type 2 objects (class='K') are included for user reference at $\approx$90\% completeness from the literature, and are generally taken as narrow-line emitters although legacy usage varied.  Also added are a selection of objects from the large SDSS releases which did not classify type 2 objects but included emitters not subclassed as ``BROADLINE'', many of which will be type-2 objects.  I used rules to include them as follows:  firstly, 1264 good confidence pipeline-only quasars not flagged as ``BROADLINE'' were taken.  Also, 1089 manual quasars with those same attributes were taken, because the manual catalogues included but did not classify type-2 objects, see P\^{a}ris et al. \shortcite{DR9Q}, introduction, last paragraph.  Next, 16842 SDSS galaxies subclassed as ``AGN'' but not ``BROADLINE'' were taken as narrow line Seyferts although some will be LINERs; similarly, 3247 SDSS galaxies subclassed as ``AGN BROADLINE'' were taken up as AGN (class='A').  These 20089 galaxy sourced objects are flagged in the HMQ comment field with an 'a' to signal their provenance.  In this way the germane active objects from the SDSS releases have been included, although the HMQ class='K' consequently contains unquantified contamination from broad-line quasars and LINERs.    

BL Lac objects (class='B') have been quality checked and ones without radio/X-ray association which VCV annotated as ``BL?'' have been dropped.  Newly surveyed such objects are usually presented as ``candidates'', so I've added those where confirmed by radio/Xray association.  There is an on-line compendium of blazars, the BZCAT \cite{MASSARO} which has been used as a check and a reference, but not as a definitive guide.  

The AGN (class='A') are in principle broadline Seyfert galaxies which are optically dominated by their disk, with QSOs (class='Q') being core-dominated objects with no disk seen.  Of course the actual data shows intermediate objects throughout the full range, in line with the unified model.  No attempt is made to subclass the AGN as S1.2, S1.5, etc.  VCV used a cosmology-based luminosity threshold to separate the AGN and QSOs, and re-released their final edition with a changed cosmology.  By contrast, the approach here is simply a practical one to separate the visually extended AGN from the unresolved QSOs, analyze those groups to get a luminosity based divider, and extend that divider into the set of fainter unresolved objects.  If done correctly, the separator should trace a path from the bright QSO-AGN divider to some natural break in the bulk fainter data profile as is seen in Figure 3.  With some testing I've identified a simple luminosity equation: 

\begin{center}
\textsl{magnitude = 22 + 5 $\times$ log$_{10}$( z $\times$ (1 + z ))} 
\end{center}

\noindent which traces the required path.  However, type 1 objects of z$<$0.1 (as a rough cut) are all taken as AGN Seyferts, consistent with the absence of nearby quasars.  Thus I need to adjust the separator to terminate at z=0.1.  Doing so gives: 

\begin{center}
\textsl{magnitude = 22 + 5 $\times$ log$_{10}$( z $\times$ (0.8 + z ) - 0.09)} 
\end{center}
  
\noindent as the separator which is bounded by Vmag=17.0 at z=0.1.  However, such bright objects can be classified by their psf.  The separator is needed for the set of objects which are psf-unresolved just because they are so faint; these are SDSS objects with g$\geq$22.0 (``survey-quality threshold'', P\^{a}ris et al. 2014, section 2.3) and vmag$\geq$21.0 for other objects.  The separator is also needed for brighter objects which have no psf class, e.g., USNO-B photometry.  This luminosity-based separator equation is that of a static Euclidean manifold with time dilation, so is non-physical in a scale-invariant universe, but fits the data well.  Figure 3 shows the separator at \textsl{x=0} as a local minimum in the data profile of the faint objects.  The QSOs to the left (brighter) of the separator comprise the bulk of the objects; the twin peaks are an artefact of SDSS BOSS\footnote{Baryon Oscillation Spectroscopic Survey, http://www.sdss3.org/surveys/boss.php} selection.  The AGN to the right (fainter) of the separator have a Gaussian-looking profile which, if real, implies some underlying physical distinction between the QSOs and AGN.       

\begin{figure}[h] 
\includegraphics[scale=0.375, angle=0]{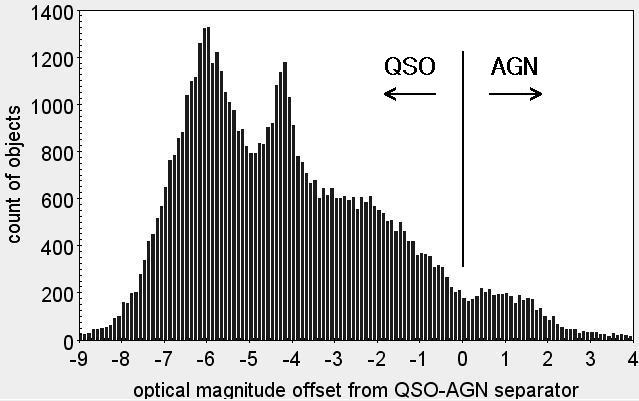} 
\tiny
{Chart produced with TOPCAT \cite{TAYLOR}.}  
\caption{Vmag offsets from luminosity separator for faint (see text) or no-psf type-1 objects, binned by 0.1 mag} 
\end{figure}

\noindent In summary of type 1 objects, extended psf morphology signals an AGN, with SDSS morphology used for SDSS photometry.  Galaxies from the Principal Galaxy Catalogue \cite{PGC} are taken as extended, as well as any type-1 of z$<$0.1.  Brighter unresolved objects (as defined above) are classified as QSO, and fainter unresolved objects, and those without psf class, are classified as QSO or AGN using the above separator equation; the overall outcome is $\approx$94\% QSOs.

\section{Optical photometry}

The HMQ sources optical photometry from a background data pool as is documented in Flesch \& Hardcastle \shortcite{FH}, but some points are worth highlighting.  The optical data is presented as red \& blue psf and magnitudes, and are preferentially sourced from Cambridge Automatic Plate Measuring machine (APM) \cite{MI} and United States Naval Observatory \cite{MONET} USNO-B data.  In particular, their data from the first-epoch National Geographic-Palomar Observatory Sky Survey (POSS-I) are preferred because POSS-I $O$ is violet 4100\AA\ and so well-offset from POSS-I $E$ 6500\AA\, and both plates were always taken on the same night, thus the red-blue colour is correct even for variable objects.  Other surveys' photometry are used where POSS-I data is incomplete, and SDSS photometry is used where both red and blue magnitudes are not already available from the APM or USNO-B.  The HMQ ReadMe has additional information about the photometry and how it is flagged in the data.

\section{Radio and X-ray associations}

HMQ displays radio/X-ray associations just by giving the radio/X-ray identifier(s) without elaboration.  Beneath this simple presentation are the calculations detailed in Flesch \& Hardcastle \shortcite{FH} which quantify causality, i.e., the confidence that the association is true.  In brief, the method is that areal densities of optical profiles around radio/X-ray detections are compared to background averages to obtain the causality percentage.  These calculations are done against anonymous optical data, and catalogues like the Atlas of Radio/X-ray Associations \cite{FL10} which report associations to anonymous optical objects do need to display those calculated probabilities.  But when that anonymous object turns out to be a quasar, which are commonly radio/X-ray emitters, then that's a strong confirmation that the association is correct.  Thus, since HMQ has quasars only, this very simple presentation style can be used.    

The radio/X-ray associations were useful in making many identifications, e.g., the quasar ``Q 1224-116'' is identified as the \textit{r=18.3 b=18.6} object at 12 27 25.0 -11 33 39 (J2000) because of the X-ray association 3XMM J122725.1-113340, whereas previously it was identified to the nearby galaxy 2MASSi J1227247-113334; the original author did not distinguish between them.  This theme was encountered in many places.  

The ReadMe itemizes the radio/X-ray surveys used, but of course they are all the largest ones like the NRAO VLA Sky Survey catalog (NVSS: Condon et al. 1998), the Faint Images of the Radio Sky at Twenty-cm survey catalog (FIRST: White et al. 1997), and the Sydney University Molonglo Sky Survey \cite{SUMSS} for the radio data, and \textit{Chandra}, \textit{XMM-Newton}, \textit{Swift}, and \textit{ROSAT} catalogs for the X-ray data.  Optical field solutions are calculated from the raw source positions of all these catalogs as described in Flesch \shortcite{FL10}.  Radio and X-ray associations are calculated independently of each other. 

The radio/X-ray identifiers are displayed in four columns which are one for X-ray, one for core radio, and two for double radio lobes.  Some association identifications are seen in the sample data of Table 1, and Table 8 shows the total HMQ count of association types by object class.  However, radio lobes are uncommon ($<$1\%), so when there are no lobes those two columns are instead used to display additional radio/X-ray identifiers from other surveys; this conveys a fuller sense of the data available.  The ReadMe gives the rules used.  The four columns are usually enough to display all available associations, but any overflow is not shown; this happens for 2891 objects with a total of 4241 X-ray associations dropped, representing 5.7\% of all X-ray associations.  An example is the blazar ``ON 231'', seen in Table 1 with 4 associations but has additional X-ray associations 2RXP J122132.0+281403, 1WGA J1221.5+2813 and 1RXS J122131.1+281402.  Such extra identifiers would be nice to show but it's not essential because they all refer to the same physical X-ray source seen on the sky, so are duplicates in that sense.

\begin{table}[h]
\scriptsize	 
\caption{Counts of Radio/X-ray associations in the HMQ}
\begin{tabular}{lrrrrrr}
\hline 
object      & total  & \#    & core  & double & extra & extra \\
class      & \# objs & X-ray & radio & lobes  & X-ray & radio \\
\hline
QSOs        & 424748 & 29086 & 21955 & 1652   & 9774  & 10538 \\
AGN         & 26623  & 6194  & 5394  & 256    & 2904  & 2283  \\
photometric & 25015  & 16954 & 8336  & 438    & 4281  & 4285  \\
BL Lac type & 1595   & 995   & 1430  & 13     & 605   & 767   \\
type 2      & 32783  & 2419  & 4697  & 271    & 860   & 1822  \\
\hline
ALL         & 510764 & 55648 & 41812 & 2630   & 18424 & 19695 \\
\hline
\end{tabular}
\end{table}

\section{Inclusion of photometric quasars}

This catalogue includes 25015 high-confidence (99\%) SDSS photometric quasar candidates which also show radio and/or X-ray associations.  These objects are well assured of being true quasars because two independent procedures, the photometry calculations and the radio/X-ray association calculations, separately identify the same objects as likely quasars, and the calibrated joint calculation given below quantifies that to the 99\% likelihood required.  

These objects are within the scope of this catalogue because the confidence in the spectroscopically confirmed quasars, i.e., that they are true quasars, are also rated at about 99\%.  This assessment is based on historical performance, and three examples convey the point.  First, the VCV quasar catalogue, heretofore the authoritative standard, was found above to have $\approx$5000 galaxies or stars amongst its 168941 nominal active objects for an error rate of 3\% and thus performance of 97\%.  Those have been removed but an unquantified residue remains from legacy surveys.  The second example is that of the 20298 SDSS pipeline-only quasars, documented above, which are included because the preponderance of them are true quasars, but some noisy spectra are present which conceal an error rate which I can estimate only informally at about 5\%.  Thirdly, even the high-quality SDSS manual surveys have a typical drop rate of 0.15\% per each new edition (Schneider et al. 2010, section 6), and in the case of DR10Q which presented 169459 QSOs, the subsequent DR12Q release reclassified 1655 of those into galaxies and dropped 25 entirely, for a total 0.99\% drop rate.  So false positives slip through even under state-of-the-art conditions.  Thus, over the set of all classified quasars in the literature, my assessment of an overall 1\% error rate is a fair call.

Accordingly, the confirmed objects can be supplemented with photometric candidates which as a group also have 99\% confidence of being true quasars.  These photometric quasars are from the SDSS-based NBCKDE (i.e., Nonparametric Bayes classifier - kernel density estimate) \cite{NBCKDE} and XDQSO (i.e., Extreme-deconvolution QSO targets) \cite{XDQSO} photometric quasar candidate catalogues.  These catalogues give nominal estimates of pQSO (probability of being a true QSO) for each object which however are not calibrated against confirmed classified objects.  I have performed that calibration against DR12Q classifications, and the details of that calibration and discussions of the NBCKDE and XDQSO attributes are in Appendix A.  This yields a calibrated photometrically-based QSO likelihood for each candidate.

Next, QSO likelihoods are needed for the radio/X-ray associations.  The calculations described in section 7 obtain the confidence of association percentage, i.e., causality, and the classified objects used in that analysis include a percentage of QSOs; thus, causality $\times$ QSO \% = QSO likelihood.  This likelihood is already calibrated against confirmed objects as part of the calculation.  Double radio lobes are included via a heuristic algorithm, also calibrated.

Now the calibrated photometric based QSO likelihood P$_{1}$ and the radio/X-ray association based QSO likelihood P$_{2}$ are combined as (expressing the likelihoods as fractions): 

\begin{center}
\textsl{P$_{tot}$ = 1 / (1 + ((1 - P$_{1}$)(1 - P$_{2}$) / P$_{1}$P$_{2}$ ))}
\end{center}

\noindent This equation requires a component to be above 50\% for it to make a positive contribution to the outcome; therefore we consider only objects with both P$_{1}>$50\% and P$_{2}>$50\%.  The goal is to identify the set of candidates with P$_{tot}\geq$99\%.  Table 9 shows the result, binned by rounded P$_{tot}$, and the three right-hand columns show the results for fellow candidates which have since been classified by SDSS-DR12Q (with ZWARNING=0 for best precision).  Column 6 gives the actual QSO percentage of classified objects, and shows a performance break between the rows binned at 96\% and 97\%, so I provisionally accept the bins of 97\% to 100\% (i.e., the bottom 4 rows) as equivalent actual performers, and sum them at bottom.  For that total set, the column 3 calculated likelihood is 98.977\% which is acceptably close to the 99\% criterion, and the column 6 actual QSO percentage of 99.175\% meets the criterion.  In this way the 25015 are selected.

\begin{table}[h] 
\scriptsize	 
\caption{QSO likelihoods for the SDSS candidates, with confirmed percentages}
\begin{tabular}{c@{\hspace{2pt}}|r@{\hspace{4pt}}c@{\hspace{2pt}}|
                r@{\hspace{2pt}}r@{\hspace{4pt}}c@{\hspace{2pt}}}
P$_{tot}$ & candidate & P$_{tot}$ &  DR12Q  &  DR12Q  &  DR12Q  \\ 
bin     & \# objects  & avg pct   & \# objs & \# QSOs &  QSOpct \\ 
\hline 
  90\%  &      403    &  90.024   &     63  &     62  &  98.413 \\
  91\%  &      492    &  91.018   &     85  &     82  &  96.471 \\
  92\%  &      580    &  92.038   &    102  &     99  &  97.059 \\
  93\%  &      703    &  93.041   &    159  &    155  &  97.484 \\
  94\%  &      873    &  94.031   &    162  &    160  &  98.765 \\
  95\%  &     1164    &  95.035   &    225  &    217  &  96.444 \\
  96\%  &     1575    &  96.030   &    334  &    326  &  97.605 \\
  97\%  &     2464    &  97.044   &    475  &    471  &  99.158 \\
  98\%  &     4267    &  98.055   &    840  &    828  &  98.571 \\
  99\%  &     8286    &  99.065   &   2069  &   2059  &  99.517 \\
 100\%  &     9998    &  99.774   &   3041  &   3014  &  99.112 \\
\hline
97-100  &    25015    &  98.977   &   6425  &   6372  &  99.175 \\
\hline
\end{tabular} 
\end{table} 

\noindent These 25015 photometric objects are given class='q' in this catalogue to identify them as objects without full spectra, but photometric redshifts are displayed.  These are sourced as available from NBCKDE which provides redshifts for 14202 NBCKDE-only and shared objects.  The remaining 10813 objects are XDQSO-only, but XDQSO has not published photometric redshifts.  These I have calculated via a clustering method akin to rainflow analysis which is described in Appendix B.  Note that photometric redshift calculations sometimes fail to achieve a result, for example when two contending redshift values are adjudged equally likely.  This was the case for an additional 627 otherwise-qualifying candidates, representing a failure rate of 2.4\%, which were accordingly excluded from the HMQ because redshifts are required.  All photometric redshifts are rounded to 0.1z in the catalogue to help convey that they are estimates.  Sample objects are seen at top \& bottom rows of Table 1.

\section{Miscellaneous Notes}

The HMQ includes 6770 new spectroscopically confirmed QSOs (and 4560 type-2) from 101 papers published since the VCV 13$^{th}$ edition, apart from the SDSS surveys.  My thanks to those authors who sent me lists of their objects.  Also, 180 QSOs from 67 early papers overlooked by VCV have also been added.  

The biggest ongoing task throughout this project was de-duplication of objects across papers and surveys, while preserving true doublets.  These are all resolved except maybe for two BL Lac objects (or is it one) which are SHBL J23435+3440 at J234333.8+344005 and BZBJ2343+3439 at J234333.5+343948.  These objects are 14 arcsec apart, and both have redshift=0.336.  The SHBL object has radio NVSS J234333.7+344000 and the BZB object has X-ray 1SXPS J234333.4+343949.  The BZB object has white BL Lac colours while the SHBL object is quite red, but its discovery paper was clear on the identification.  I suspect the SHBL object is the true one and the BZB object a duplicate, but possibly both are valid so both are in this catalogue.      

The HMQ is a catalogue of quasars and not of gravitational lens images, so I've de-duplicated multiple lensed signatures into single identifications where I was aware of them; the optically brightest signature is the one retained.   

I write NVSS radio detections in the form \textsl{NVSS Jhhmmss.s+ddmmss} which is the standard form used by radio/X-ray surveys, although the recommended NVSS naming convention is \textsl{NVSS Jhhmmss+ddmmss}.  I prefer the standard form because when both FIRST and NVSS report the same signature, their astrometry agrees well, often to the arcsecond, so the standard form displays useful astrometric information.  The recommended NVSS name can be recovered by simply removing the time decimal.

\section{Conclusion}

The Half Million Quasars (HMQ) catalogue is presented as a comprehensive edition, complete to 25 January 2015, which displays 451371 type 1 QSOs and AGN, 25015 high-confidence photometric quasars, 1595 BL Lac objects, and 32783 type 2 objects.  Astrometry is arcsecond accurate for most objects, out to 2 arcsec precision for some legacy objects.  X-ray and radio associations for these quasars are presented as applicable, including double radio lobes.  Issues of identification and classification have been resolved to where the full quasar portfolio can be easily integrated into large databases.

\begin{acknowledgements}
Thanks to NASA HEASARC for putting an early version of this catalogue on-line which motivated me to clean up all the issues with it, although I had no idea that it would take me five years to do so.  Thanks to Dave Monet for generously providing USNO-B optical data.  Thanks to Steve Willner for finding papers which I could not, Emanuele Farina for keeping me on my toes with some early duplicates, and to Isabelle P\^{a}ris for helpful discussions.  Thanks to the sci.astro.research newsgroup, astutely moderated by Martin Hardcastle, which gave me a platform to hammer out issues and get valuable feedback.  This work was not funded.
\end{acknowledgements}

\clearpage
\appendix{\textbf{\large{APPENDICES}}} 
\section{Photometric Likelihood Calibration}

The HMQ includes 25015 photometric quasars from the SDSS-based XDQSO \cite{XDQSO} and NBCKDE \cite{NBCKDE} photometric quasar candidate catalogues.  These are a small fraction of the total objects from these catalogues.  XDQSO data consists of 160M objects in 39Gb of files, most of which are stars, and provides a nominal pQSO (percentage chance of being a QSO) which their paper cautions is just a comparative figure and not a genuine QSO likelihood (see esp. their figure 14).  The primary purpose of XDQSO was to select targets for the SDSS BOSS campaign, and those are flagged with good=0 (which means good).  Those not so flagged could still have a high pQSO but would have deficiencies such as being optically too faint, etc.  3010139 XDQSO candidates have pQSO$>$0.5 of which 1792678 were BOSS targets and 1217461 were non-BOSS.  NBCKDE's final release was as a file of 1015082 ``most robust'' candidates, with the pQSO presented as comparative QSO and star density logarithms; 992700 of these have pQSO$>$0.5.    

To get best use of the pQSOs, it is desirable to calibrate them such that they represent the true odds of the candidate being a QSO (the ``QSOpct'').  The subsequent SDSS BOSS releases DR9-DR12 were of course largely onto targets selected by XDQSO and NBCKDE, so a large pool of spectroscopically classified objects are available to reconcile to; the SDSS-DR12Q \cite{DR12Q} superset has 531648 manually classified quasars, stars, and galaxies, of which 331188 were XDQSO BOSS targets, 170145 were NBCKDE candidates, and 7289 were XDQSO non-BOSS objects.  However, there is extensive overlap between XDQSO and NBCKDE, and in the calibration exercise it is desirable to separate them lest objects be double-counted which would denormalize the calibration.  Germane to this is that the NBCKDE footprint (DR6 based) is completely within the XDQSO footprint (DR8 based), so removal of the shared objects will still allow NBCKDE to be treated as an internally consistent dataset.  Therefore the 331188 XDQSO BOSS targets are processed as the first priority, while just 24127 NBCKDE candidates not matching to BOSS remain as the second group to be processed.  However, the performance of the complete NBCKDE data is interesting in its own right, and will be analyzed below also.

Of the 7289 non-BOSS XDQSO objects classified by DR12Q, 4169 are shared with NBCKDE but are left to be processed with the NBCKDE group.  This is because the vast majority of non-BOSS objects are optically very faint or have photometry issues, and the subset classified by DR12Q were picked from the more eligible members, thus not very representative of the whole group.  Exclusion of the NBCKDE-shared objects lowers the QSO percentage of the remaining 3120 non-BOSS candidates more to what is expected for the group as a whole.  Thus the 3120 non-BOSS XDQSO objects which are not shared with NBCKDE are the final group to be processed.  

The calibration procedure is presented as two tables, and each table presents figures for four groups of candidates, being (1) XDQSO BOSS, (2) NBCKDE full set, (3) NBCKDE-only candidates (not shared with BOSS), and (4) XDQSO non-BOSS candidates which are not shared with NBCKDE.  Table A1 shows the performance of the nominal pQSOs for each group when compared to the DR12Q-revealed classifications.  The pQSOs are binned by 5\% in the range 60\% to 85\% (with the displayed figure being the midpoint of the bin), then binned by 1\% in the range 88\% to 100\%; this highlights the high percentage bins which have the most numbers and the best performance.  

For each group, the ``DR12Q QSOpct'' column is the calibrator; we want a mapping which will convert the pQSO (i.e., the nominal QSO expectancy) to that QSOpct (i.e., the calibrated actual performance), but it needs to be a simple mapping to smooth the outcome.  I've used these simple mappings to achieve the calibration:

\begin{enumerate}
	\item XDQSO BOSS targets: \\
	      for pQSO $\leq$ 90: QSOpct = pQSO - 14 \\
	      for 90 $<$ pQSO $\leq$ 96: QSOpct = 2 $\times$ pQSO - 104 \\
	      for pQSO $>$ 96: QSOpct = 2.5 $\times$ pQSO - 152
	\item NBCKDE full data: \\
	      for pQSO $\leq$ 90: QSOpct = pQSO - 12 \\
	      for pQSO $>$ 90: QSOpct = 2 $\times$ pQSO - 102
	\item NBCKDE-only data: \\ 
	      for pQSO $\leq$ 92: QSOpct = pQSO - 36 \\
	      for pQSO $>$ 92: QSOpct = 5 $\times$ pQSO - 404
	\item XDQSO non-BOSS data: \\
	      QSOpct = 30 + pQSO/2
\end{enumerate}

\begin{table*}
\scriptsize	 
\caption{Before Calibration: XDQSO/NBCKDE pQSOs compared to DR12Q superset QSOpcts}
\begin{tabular}{@{\hspace{0pt}}c|@{\hspace{2pt}}r@{\hspace{4pt}}r@{\hspace{4pt}}c@{\hspace{4pt}}c|@{\hspace{2pt}}r@{\hspace{4pt}}r@{\hspace{4pt}}c@{\hspace{4pt}}c|@{\hspace{2pt}}r@{\hspace{4pt}}r@{\hspace{4pt}}c@{\hspace{4pt}}c|@{\hspace{2pt}}r@{\hspace{4pt}}r@{\hspace{4pt}}c@{\hspace{4pt}}c}
\hline 
 & \multicolumn{4}{c}{XDQSO BOSS targets} & \multicolumn{4}{c}{NBCKDE full data} 
 & \multicolumn{4}{c}{NBCKDE-only data$^{*}$} & \multicolumn{4}{c}{XDQSO non-BOSS$^{\dagger}$} \\ 
pQSO & DR12Q  & DR12Q  & DR12Q  & calib & DR12Q & DR12Q  & DR12Q  & calib & DR12Q & DR12Q  & DR12Q  & calib & DR12Q & DR12Q  & DR12Q  & calib \\
 bin & \#objs & \#QSOs & QSOpct & bin & \#objs & \#QSOs & QSOpct & bin & \#objs & \#QSOs & QSOpct & bin & \#objs & \#QSOs & QSOpct & bin \\
\hline
 60  & 16063 & 7963  & 49.6 & 46   & 1227  & 848   & 69.1 & 48   & 423  & 215  & 50.8 & 24   & 233 & 145 & 62.2 & 60 \\
 65  & 16571 & 8813  & 53.2 & 51   & 4164  & 2264  & 54.4 & 53   & 1849 & 707  & 38.2 & 29   & 204 & 126 & 61.8 & 62.5 \\
 70  & 17223 & 9793  & 56.9 & 56   & 6190  & 3612  & 58.4 & 58   & 2353 & 932  & 39.6 & 34   & 206 & 142 & 68.9 & 65 \\
 75  & 18405 & 11202 & 60.9 & 61   & 7358  & 4647  & 63.2 & 63   & 2213 & 860  & 38.9 & 39   & 185 & 130 & 70.3 & 67.5 \\
 80  & 20251 & 13176 & 65.1 & 66   & 10423 & 7054  & 67.7 & 68   & 2589 & 1139 & 44.0 & 44   & 216 & 154 & 71.3 & 70 \\
 85  & 22903 & 16171 & 70.6 & 71   & 13157 & 9426  & 71.6 & 73   & 2359 & 1091 & 46.2 & 49   & 193 & 145 & 75.1 & 72.5 \\
\hline
 88  & 5147  & 3771  & 73.3 & 74   & 3119  & 2341  & 75.1 & 76   & 479  & 255  & 53.2 & 52   & 48  & 37  & 77.1 & 74 \\
 89  & 5500  & 4116  & 74.8 & 75   & 3501  & 2644  & 75.5 & 77   & 518  & 238  & 45.9 & 53   & 44  & 37  & 84.1 & 74.5 \\
 90  & 5780  & 4394  & 76.0 & 76   & 3924  & 3090  & 78.7 & 78   & 502  & 267  & 53.2 & 54   & 36  & 25  & 69.4 & 75 \\
 91  & 6082  & 4791  & 78.8 & 78   & 4446  & 3532  & 79.4 & 80   & 464  & 263  & 56.7 & 55   & 39  & 31  & 79.5 & 75.5 \\
 92  & 6628  & 5235  & 79.0 & 80   & 5108  & 4150  & 81.2 & 82   & 516  & 285  & 55.2 & 56   & 31  & 22  & 71.0 & 76 \\
 93  & 7355  & 6005  & 81.6 & 82   & 6606  & 5520  & 83.6 & 84   & 603  & 356  & 59.0 & 61   & 46  & 38  & 82.6 & 76.5 \\
 94  & 8145  & 6804  & 83.5 & 84   & 9334  & 7983  & 85.5 & 86   & 732  & 463  & 63.3 & 66   & 45  & 35  & 77.8 & 77 \\
 95  & 9303  & 7959  & 85.6 & 86   & 13782 & 12258 & 88.9 & 88   & 832  & 605  & 72.7 & 71   & 44  & 31  & 70.5 & 77.5 \\
 96  & 11022 & 9714  & 88.1 & 88   & 16565 & 15084 & 91.1 & 90   & 983  & 774  & 78.7 & 76   & 62  & 46  & 74.2 & 78 \\
 97  & 13424 & 12085 & 90.0 & 90.5 & 16749 & 15477 & 92.4 & 92   & 1100 & 923  & 83.9 & 81   & 49  & 37  & 75.5 & 78.5 \\
 98  & 17977 & 16742 & 93.1 & 93   & 17426 & 16251 & 93.3 & 94   & 1536 & 1337 & 87.0 & 86   & 92  & 70  & 76.1 & 79 \\
 99  & 29662 & 28319 & 95.5 & 95.5 & 15323 & 14628 & 95.5 & 96   & 1767 & 1624 & 91.9 & 91   & 119 & 96  & 80.7 & 79.5 \\
100(hw) & 38961 & 37452 & 96.1 & 97.4 & 9056 & 8741 & 96.5 & 97.5 & 1141 & 1058 & 92.7 & 94.8 & 197 & 101 & 51.3 & 79.9 \\
\hline
ALL  & 331188 & 238470 & 72.0 &    & 170145 & 141240 & 83.0 &    & 24127 & 13959 & 57.9 &    & 3120 & 2009 & 64.4 &  \\
\hline
\multicolumn{17}{l}{$^{*}$data not shared with XDQSO BOSS} \\
\multicolumn{17}{l}{$^{\dagger}$data not shared with NBCKDE} \\
\end{tabular}
\end{table*}

These mappings converts the pQSO bin to the calibrated QSOpct value displayed in the column ``calib bin''.  Table A2 shows this same data when binned by these calibrated QSOpcts, so we can better inspect the outcome.  The smoothed QSOpct bin at left should ideally match the ``DR12Q QSOpct'' column value in each of the four groups of candidates; in practice, the XDQSO BOSS and full NBCKDE data are well-fit, the NBCKDE-only data is adequately fit, and the XDQSO non-BOSS data is only marginally fit but the numbers are small.  These tables don't show the low-pQSO bins where the numbers are small and the calibrations are ill-fitting and not useful, but those poorly performing objects are not of interest to this exercise.  The calibration shows the effect of removing the XDQSO-shared objects from the NBCKDE data: the NBCKDE-only pQSOs are about 15\% less well performed than the full NBCKDE pQSOs, showing that the shared objects were the best targets.  For the full catalogues, the NBCKDE and XDQSO BOSS pQSOs have similar performance, NBCKDE being very slightly the better performed.   

The full listing of 358435 SDSS-DR12Q superset classified objects and the XDQSO and NBCKDE pQSOs for them is available for inspection\footnote{at http://quasars.org/photometric-DR12Q.txt (25Mb file)}.  Thus in this way, the XDQSO and NBCKDE pQSOs are calibrated into true QSOpcts, thus enabling the calculations of section 8 which yield the additional 25015 objects for the HMQ catalogue.

\begin{table*}
\scriptsize	 
\caption{After Calibration: XDQSO/NBCKDE QSOpcts compared to DR12Q superset QSOpcts}
\begin{tabular}{@{\hspace{0pt}}c|rrc|rrc|rrc|rrc}
\hline 
 & \multicolumn{3}{c}{XDQSO BOSS targets} & \multicolumn{3}{c}{NBCKDE full data} 
 & \multicolumn{3}{c}{NBCKDE-only data$^{*}$} & \multicolumn{3}{c}{XDQSO non-BOSS$^{\dagger}$} \\ 
QSOpct & DR12Q  & DR12Q  & DR12Q  & DR12Q  & DR12Q  & DR12Q  & DR12Q  & DR12Q  & DR12Q  & DR12Q  & DR12Q  & DR12Q  \\
  bin  & \#objs & \#QSOs & QSOpct & \#objs & \#QSOs & QSOpct & \#objs & \#QSOs & QSOpct & \#objs & \#QSOs & QSOpct \\
\hline
 60 & 18303 & 10945 & 59.8 & 6586  & 3977  & 60.4 & 590  & 345  & 58.5 & 441 & 275 & 62.4 \\
 65 & 19632 & 12615 & 64.3 & 8362  & 5472  & 65.4 & 716  & 440  & 61.5 & 414 & 287 & 69.3 \\
 70 & 22268 & 15482 & 69.5 & 11406 & 7850  & 68.8 & 820  & 599  & 73.0 & 398 & 280 & 70.4 \\
 75 & 22843 & 17007 & 74.5 & 14994 & 11061 & 73.8 & 976  & 770  & 78.9 & 405 & 315 & 77.8 \\
 80 & 16698 & 13333 & 79.8 & 12333 & 9848  & 79.9 & 1079 & 892  & 82.7 & 537 & 362 & 67.4 \\
 85 & 21859 & 18489 & 84.6 & 20720 & 17617 & 85.0 & 1483 & 1278 & 86.2 & 0   & 0   &   \\
\hline
 88 & 4909  & 4338  & 88.4 & 6951  & 6196  & 89.1 & 175  & 162  & 92.6 &     &     &   \\
 89 & 4835  & 4313  & 89.2 & 8046  & 7273  & 90.4 & 312  & 286  & 91.7 &     &     &   \\
 90 & 5022  & 4496  & 89.5 & 8397  & 7654  & 91.2 & 341  & 304  & 89.1 &     &     &   \\
 91 & 5617  & 5079  & 90.4 & 8177  & 7508  & 91.8 & 559  & 521  & 93.2 &     &     &   \\
 92 & 6228  & 5688  & 91.3 & 8401  & 7758  & 92.3 & 364  & 335  & 92.0 &     &     &   \\
 93 & 7160  & 6682  & 93.3 & 8677  & 8034  & 92.6 & 191  & 178  & 93.2 &     &     &   \\
 94 & 8383  & 7900  & 94.2 & 8887  & 8297  & 93.4 & 379  & 358  & 94.5 &     &     &   \\
 95 & 10296 & 9784  & 95.0 & 8022  & 7573  & 94.4 & 373  & 348  & 93.3 &     &     &   \\
 96 & 13136 & 12569 & 95.7 & 7677  & 7350  & 95.7 & 389  & 352  & 90.5 &     &     &   \\
 97 & 18040 & 17431 & 96.6 & 7382  & 7122  & 96.5 & 0    & 0    &      &     &     &   \\
 98 & 24873 & 23835 & 95.8 & 4649  & 4484  & 96.5 &      &      &      &     &     &   \\
 99 & 0     & 0     &      & 0     & 0     &      &      &      &      &     &     &   \\
\hline
\multicolumn{13}{l}{$^{*}$data not shared with XDQSO BOSS} \\
\multicolumn{13}{l}{$^{\dagger}$data not shared with NBCKDE} \\
\end{tabular}
\end{table*}

\appendix
\section{Photometric Redshifts made easy}

Statistical tasks can be made easy if a large training set is available and the data is well defined.  So it was for me when I first calculated photometric redshifts for the XDQSO BOSS candidates, because the SDSS-DR9 provided 199751 quasar redshifts as a training set.  I have now recalculated these using 386303 quasar redshifts from the DR12Q manual catalogue, which of course yields little change since the DR9 training set was already very large.

SDSS photometry is of 5 bands known as \textsl{ugriz} with midpoints at approx 3600\AA, 4800\AA, 6200\AA, 7600\AA, and 9000\AA, respectively.  These magnitudes provide 4 colours: \textsl{ug} (=\textsl{u-g}), \textsl{gr}, \textsl{ri}, and \textsl{iz}.  My procedure is to class each candidate by those colours (rounded to 0.1 mag), and the training set provides spectroscopic redshifts for all classified quasars having those same colours.  So for a candidate having colours \{\textsl{ug=0.2 gr=0.2 ri=-0.1 iz=0.4}\}, the profile of spectroscopically confirmed redshifts is shown in Figure B1.

\begin{figure}[h] 
\begin{center} 
\includegraphics[scale=0.4, angle=0]{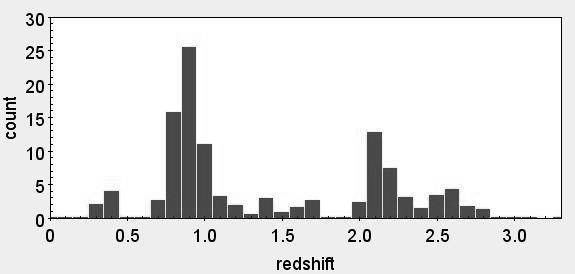} 
\caption{count of spectroscopically confirmed redshifts for the colours \textsl{ug}=0.2 \textsl{gr}=0.2 \textsl{ri}=-0.1 \textsl{iz}=0.4, binned by 0.1z.} 
\end{center} 
\end{figure}
   
The colour-fitting training set objects are supplemented by near-colour fits (e.g., perturbing each colour in turn by 0.1 mag) which are weighted much less, so that candidates with unusual colour profiles will still have an adequate pool of training set objects.  The task now is to reduce this redshift profile into a single photometric redshift value -- basically, the redshift value with the greatest expectancy of success which we arbitrarily define as being within 0.5z of the true spectroscopic redshift; however, closer precision is always a goal.  The method is to cluster the redshift bins, identify the largest cluster, and select the redshift from that cluster. 

NBCKDE \cite{NBCKDE} used a kernel density estimate to cluster their training set data, and selected the "most likely" photometric redshift as the weighted average of the bounded maximal cluster for each candidate.  In this work, for simplicity, I use a variant of the ``rainflow'' method which is used in engineering to cluster a continuous record (usually of amplitudes $\times$ time) into a series of discrete events.  Rainflow analysis requires both a peak maximum and peak minimum to define an event; it is a form of profile reduction.  I apply such an algorithm to the redshift profile of Figure B1; the result is seen on Figure B2 upper left, where the 4 vertical dividers delineate the 5 clusters identified by the rainflow algorithm.  The clusters are summated and the 2nd cluster centred on z=0.9 is identified as the largest, and its largest single bin (of width 0.1z) is selected as the photometric redshift; thus the rainflow value is z=0.9.  The NBCKDE photometric redshift for this object is 0.845, the difference being from different cluster boundaries and that NBCKDE used the weighted average of the cluster; I prefer to use the maximum bin value in view of that cluster boundaries are not always clear.  This object is identified in Figure B2 as SDSS J000922.96+155201.8 and it is a DR12Q object (as all of these are) with a spectral redshift of 1.012.

\begin{figure} 
\begin{center} 
\includegraphics[scale=0.3, angle=0]{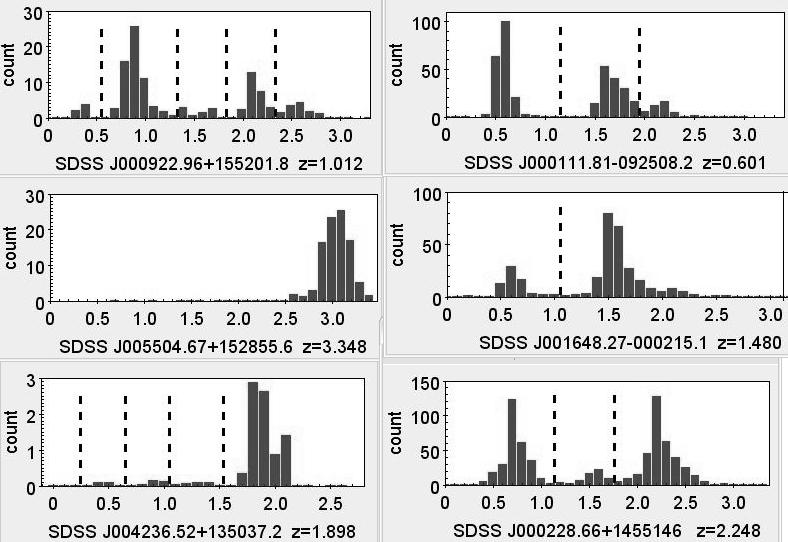} 
\caption{Six redshift profiles for photometric redshift calculations} 
\end{center} 
\end{figure}

Figure B2 shows a brief sample of redshift profiles encountered in this task.  The upper-right profile is for objects of colours \textsl{ug=0.4 gr=0.0 ri=0.3 iz=0.1} and shows 3 clusters; z=0.6 is selected as the photometric redshift.  NBCKDE selected z=1.625, evidently because they treated the grouping on the right as a single cluster, whereas the rainflow algorithm divided it.  That the spectral redshift is 0.601 is of no moment; quasars of the same colours with spectral redshift of 1.6 or 1.7 are trivially found.

Figure B2 middle-left has colours \textsl{ug=1.8 gr=0.2 ri=0.1 iz=0.0} and is obviously a high-redshift object where the Lyman break enters the optical domain and dims the \textsl{u} magnitude.  No clustering to be done, so my photometric redshift is 3.1, NBCKDE has 3.045, spectral is 3.348.

Figure B2 middle-right has colours \textsl{ug=0.4 gr=0.2 ri=0.2 iz=0.0} and shows two obvious clusters.  My photometric redshift is 1.5, NBCKDE has 1.465, spectral is 1.480.  

Figure B2 lower left has colours \textsl{ug=-0.1 gr=-0.2 ri=0.3 iz=0.1} and shows five clusters of which the one on the right is obviously the ``winner''.  My photometric redshift is 1.8, NBCKDE has 2.025, spectral is 1.898.  The low counts (vertical axis) show that this colour profile is unusual, and the fractional counts show that the training set was supplemented by perturbed values having small weights.  

Figure B2 lower right has colours \textsl{ug=0.4 gr=0.1 ri=0.1 iz=0.1} and shows three clusters of which the left and right clusters are similar in size; summations show that the 3 clusters hold 41\%, 8\%, and 51\% respectively.  My photometric redshift is 2.2, NBCKDE has 2.215, spectral is 2.248.  This redshift profile came close to yielding no result as happens when the leading contenders are too closely matched; about 2.5\% of objects get no photometric redshift because of this.     

A comprehensive comparison of the photometric redshifts from this algorithm (based on the DR9 training set) compared with NBCKDE is available for inspection\footnote{at http://quasars.org/docs/NBCKDE-MQ-redshifts.txt}.  It lists 22919 SDSS-DR10 quasars as the test data; these quasars were first presented in DR10 so there is no overlap with the training set.  Accordingly, the true spectral SDSS-DR10 redshifts are also displayed, and the testing criterion is to treat each photometric redshift as a ``hit'' if within 0.5z of the true spectral redshift.  There it is seen that the NBCKDE hit rate is 16658/22919 = 72.7\%, and the rainflow hit rate is 16986/22919 = 74.1\%.  No doubt the NBCKDE method is better than my own simple algorithm, but I had the advantage of the very large SDSS-DR9 training set.  This good performance validates the assignation of these photometric redshifts to the XDQSO-only objects.  

\end{document}